\documentclass[preprint, 11pt]{elsarticle}

\pdfoutput=1

\usepackage{natbib}
\usepackage{amsmath}
\usepackage{amstext}
\usepackage{amsfonts}
\usepackage{amssymb}
\usepackage{color}
\usepackage{graphicx}
\usepackage{hyperref}

\usepackage{amsmath}
\usepackage{amstext}
\usepackage{amsfonts}
\usepackage{amssymb}
\usepackage{graphicx}
\usepackage{hyperref}

\graphicspath{{graphics/}}

\newcommand{\pr}{^\prime }

\newcommand{\Fig}[1]{Fig.~\ref{#1}}
\newcommand{\Eq}[1]{Eq.~(\ref{#1})}

\newcommand{\Tr}{\,\mathrm{Tr}\:}
\newcommand{\Op}{\mathcal{O}}

\newcommand{\vx}{\vec{x}}
\newcommand{\vy}{\vec{y}}
\newcommand{\vp}{\vec{p}}

\newcommand{\bq}{\bar{q}}
\newcommand{\cf}{cf.\ }

\newcommand{\twop}{two-point }
\newcommand{\threep}{three-point }
\newcommand{\fm}{~\mathrm{fm}}

\newcommand{\mev}{~\mathrm{MeV}}

\biboptions{sort&compress}
\begin{document}
\begin{frontmatter}
\title{A Stochastic Method for Computing \\ Hadronic Matrix Elements
\begin{center}
\vspace{5mm}
\includegraphics[scale=0.2]{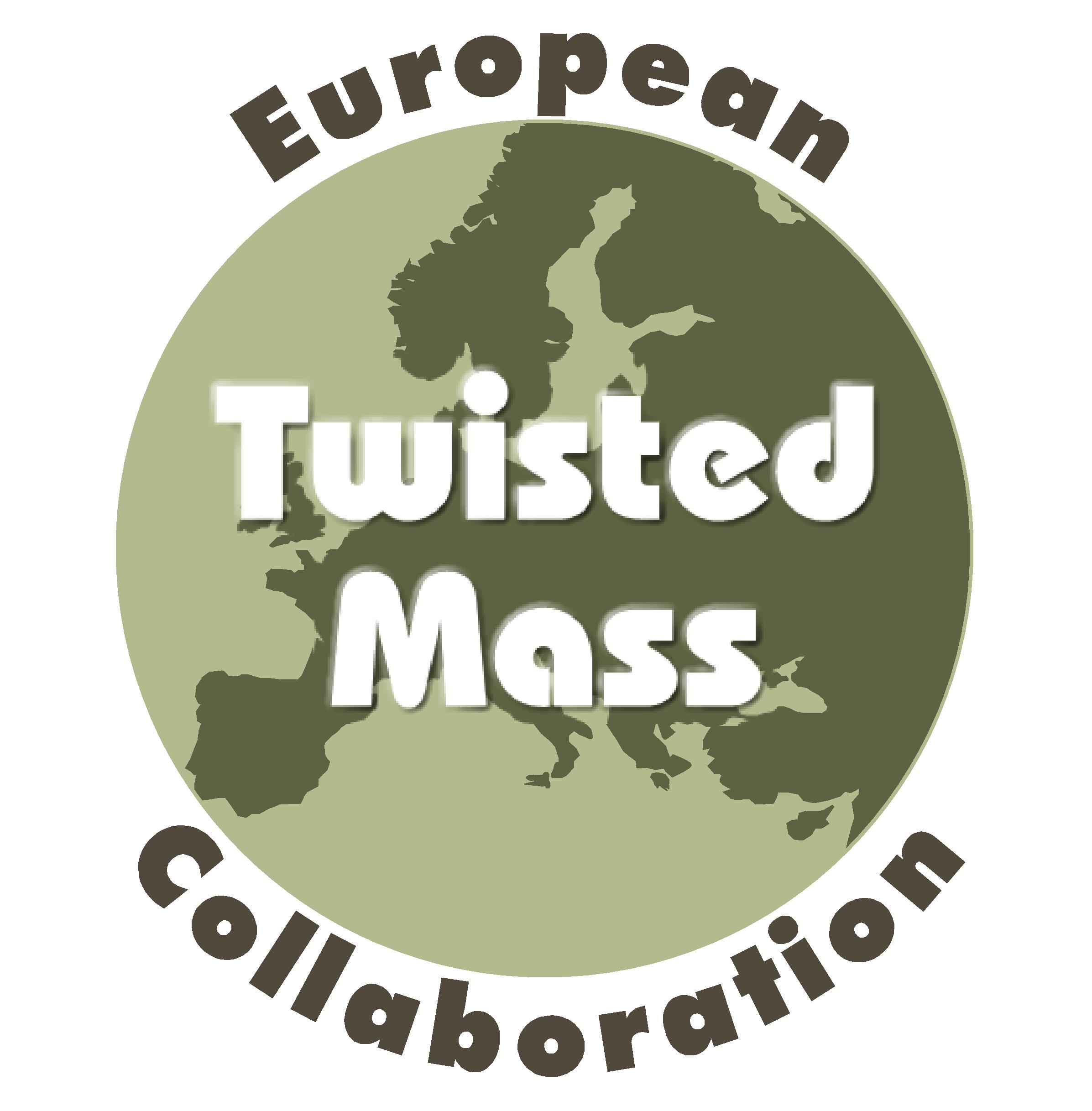}
\vspace{-5mm}
\end{center}
}

\author[UC,TCI]{Constantia Alexandrou}
\author[UC]{Martha Constantinou} 
\author[DZ]{Simon Dinter}
\author[DZ]{Vincent Drach}
\ead{vincent.drach@desy.de}
\author[DZ]{Karl Jansen}
\ead{Karl.Jansen@desy.de}
\author[UC]{Kyriakos Hadjiyiannakou}
\author[DZ,JL]{Dru B.\ Renner}

\author{\newline\\ for the ETM Collaboration}

\address[DZ] {NIC, DESY Zeuthen, Platanenallee 6, D-15738 Zeuthen, Germany}
\address[UC] {Department of Physics, University of Cyprus, P.O. Box 20537, 1678 Nicosia, Cyprus}
\address[TCI]{Computation-based Science and Technology Research Center,
              The Cyprus Institute,\\ 15 Kypranoros Str., 1645 Nicosia, Cyprus}
\address[JL] {Current address: Jefferson Lab.}

\cortext[cor1]{Corresponding author }

%
%

%
\begin{abstract}
We present a stochastic method for the calculation of baryon \threep
functions that is more versatile compared to the typically used
sequential method.  We analyze the scaling of the error of the
stochastically evaluated \threep function with the lattice volume and
find a favorable signal-to-noise ratio suggesting that our stochastic
method can be used efficiently at large volumes to compute hadronic
matrix elements.
\end{abstract}


\end{frontmatter}


\hyphenation{For-schungs-zen-trum}



%
%
\section{\label{sec:int}Introduction}
Hadron structure calculations in lattice QCD have emerged as a powerful tool
for providing comparison to and guidance
for experiments, see e.g.\ refs.~\cite{Renner:2010ks,Alexandrou:2010cm,Orginos:2011zz,Alexandrou:2012hi}.
Examples include moments of generalized parton distribution functions, as well as form factors.
Lattice computations of such quantities have been carried out at several values of
the lattice spacing allowing the continuum limit to be taken.
In addition, small, sometimes
even physical, values of the pion mass have been employed
in the calculation of these quantities
leading to an improved understanding
of their quark mass dependence and how they approach the physical point. Unfortunately,
studies 
of excited state contributions~\cite{Dinter:2011sg,Alexandrou:2012gz,Green:2011fg,Capitani:2010sg,Capitani:2012gj} suggest that for some quantities 
these effects can play an important role 
as a systematic uncertainty that affects hadronic \threep function computations.
Safely accounting for these effects requires large statistics, hence methods to speed-up these calculations
are highly desirable. 

The progress in nucleon matrix element calculations on the lattice 
has prompted an effort to go beyond the simplest observables and to pursue a larger variety of
interesting hadronic quantities.  The evaluation of more observables
will deepen our knowledge of hadron structure and provide a more
comprehensive test of 
QCD. However, using the conventional sequential 
method~\cite{Martinelli:1988rr} to calculate 
these matrix elements, it is necessary to perform 
a new computation of the needed quark propagators 
for each observable of interest\footnote{The sequential method allows to compute 
either the matrix element of a single hadron or the matrix elements of a single 
current with one sequential inversion.  
In particular, a new computation of propagators has to be performed, when computing matrix elements 
of another hadron (fixed sink method) or another current (fixed current method).}.
This then leads to a high computational 
demand if many physical quantities are being sought.

In this paper, we describe an alternative approach, based on a
stochastic method, that allows us to obtain a large class 
of observables with only a single computation of the propagators. 
To this end, we employ 
stochastically computed all-to-all propagators. Since a calculation based 
on a stochastic evaluation of propagators may
lead to very noisy results, we perform a detailed study to determine 
whether
the stochastic noise can be controlled with a moderate number
of stochastic sources. 
We determine the signal-to-noise ratio as a function of 
the lattice size to test whether our stochastic method can be used 
in large volumes, such as $48^3 \times 96$, that are used in 
contemporary lattice computations. 
These questions are addressed specifically for the example of the axial charge $g_A$ 
of the nucleon.
We would like to emphasize that we do not perform an analysis of systematic effects,
since our goal is solely to test the stochastic method.

During the course of our work, a similar stochastic approach  
was employed in the 
calculation of meson \threep functions \cite{Evans:2010tg}, 
where a (heavy) all-to-all propagator is estimated stochastically. 
There it was found that the stochastic method is competitive and 
in some cases even superior to the sequential one.
Of course, it is not guaranteed that 
the same conclusions hold for baryon matrix elements, 
since those are subject to a stronger exponentially decreasing 
signal-to-noise ratio \cite{Luscher:2010ae}.

This paper is organized as follows:\ in Sec.~\ref{sec:stoch3p} we outline our stochastic method,
in Sec.~\ref{sec:test} we present the results of this method and compare to those 
obtained by the fixed sink method, and 
we summarize our findings in Sec.~\ref{sec:sum}.
%
%
%
\section{\label{sec:stoch3p}Stochastic method for baryon three-point functions}

Quantities that are needed for 
investigating hadron structure can be extracted by 
computing matrix elements of baryons with local operators.
In lattice QCD these baryon matrix elements are obtained from 
baryon \threep functions in Euclidean space-time that are of the form
\begin{align}
 \label{eq:baryon3p}
 \sum\limits_{\vx,\vy} e^{-i\vp\pr\cdot\vx} e^{-i\vp\cdot\vy} 
\left\langle B (\vx, t) | \Op (\vy, \tau)| \bar{B} (0) \right\rangle\; , \\
\Op (\vy, \tau) = \bq(\vy, \tau) \Gamma q(\vy, \tau)\; . \label{eq:localOp}
\end{align}
$\Gamma$ represents a combination of $\gamma$-matrices and covariant derivatives and
we have used translational invariance to set the source point to zero.
Naively one would need an all-to-all propagator from every lattice point
$(\vy,\tau)$ to all points $(\vx,t)$
for the evaluation of the above \threep function, which is, of course, prohibitively 
expensive to calculate.
Such a demanding computation can be circumvented by 
applying the sequential method to perform the summation over the spatial
coordinates 
of  either the sink or the current~\cite{Martinelli:1988rr}. 
For the example of the fixed sink method, 
the momentum as well as the time slice of the sink are fixed and an additional inversion
for each flavour is needed.
An alternative approach is to estimate the all-to-all propagator stochastically,
which is the method that we explore in this paper.

A generic \threep function of a baryon $B$ is defined as
\begin{equation*}
 C_3^{(B)}\left( t, \tau ; \vp, \vp\pr \right) 
 = \zeta^{(B)}_{AA\pr} 
\sum\limits_{\vx,\vy} e^{-i\vp\cdot\vy}e^{-i\vp\pr\cdot\vx}
\left\langle  \mathcal{I}^{(B)}_{A^\prime}(\vx,t) \Op (\vy, \tau) \bar{\mathcal{I}}^{(B)}_{A} (0) \right\rangle\; .
\end{equation*}
$\mathcal{I}_{B}$ is the baryon interpolating field and $A$ and $A\pr$ 
summarize the indices depending on the quantum numbers
of the baryon $B$, which are appropriately contracted with the function $\zeta^{(B)}_{A A\pr}$. 
The insertion time of the operator is denoted by $\tau$. 
For illustration let us now consider the \threep function of a proton and the operator $\bar{d}\Gamma d$.
We use the interpolating field $\mathcal{I}_p$ possessing the quantum numbers of the proton, namely
\begin{align*}
 \mathcal{I}^{(p)}_{\alpha} (\vx,t) &= 
 \varepsilon^{abc} u_\alpha^a (\vx,t) 
 \left( \left( d^b (\vx,t) \right)^T\mathcal{C} \gamma_5 \:  u^c (\vx,t) \right)\; ,
\end{align*}
where $\mathcal{C}=i \gamma_0 \gamma_2$ is the charge conjugation operator.
In terms of quark propagators,
the connected piece of this \threep function reads
\begin{multline}
\label{eq:nucleonThreePointExample}
  C_3^{(p)}\left( t, \tau ; \vp, \vp\pr ; \mathbb{P} \right) =
  \mathbb{P}^{\alpha \alpha\pr}
  \sum\limits_{\vx,\vy}\: e^{-i\vp\cdot\vy}e^{-i\vp\pr\cdot\vx}\;
  \varepsilon_{cba}\: \varepsilon_{a\pr b\pr c\pr}\:
  \left( \mathcal{C} \gamma_5 \right)_{\beta\pr\gamma\pr}\:
  \left( \mathcal{C} \gamma_5 \right)^\ast_{\gamma\beta} \; \Gamma^{\delta\pr \delta} \delta_{d\pr d}
   \\\quad\quad\quad\quad\quad\times
   \Big[
   S^{(u)}_{(\alpha\pr a\pr) (\alpha a)} (x, 0)\,
   S^{(d)}_{(\beta\pr  b\pr) (\delta d)} (x, y)\,
   S^{(d)}_{(\delta\pr d\pr) (\beta  b)} (y, 0)\,
   S^{(u)}_{(\gamma\pr c\pr) (\gamma c)} (x, 0) \\
 - S^{(u)}_{(\alpha\pr a\pr) (\gamma c)} (x, 0)\,
   S^{(d)}_{(\beta\pr  b\pr) (\delta d)} (x, y)\,
   S^{(d)}_{(\delta\pr d\pr) (\beta  b)} (y, 0)\,
   S^{(u)}_{(\gamma\pr c\pr) (\alpha a)} (x, 0)
   \Big],
\end{multline}
where $x=(\vec{x},t)$ and $y=(\vec{y},\tau)$ and where the up (down) quark propagator is denoted by $S^{(u)}$ ($S^{(d)}$).
$\mathbb{P}$ is an appropriate spin projector, which we will specify later. 

\begin{figure}[tb]
\centering
\includegraphics[width=0.49\textwidth]{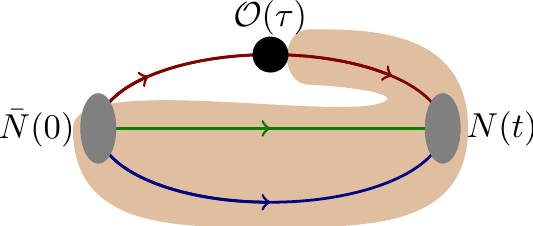}\hfill
\includegraphics[width=0.48\textwidth]{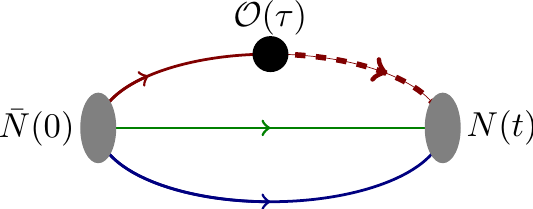}
\caption{\label{fig:diagStandardAndStochastic}
Diagrammatic illustration of the sequential method through the sink (left) 
and the stochastic method (right).}
\end{figure}

The sequential method with fixed sink makes use of the fact that we can perform 
the sum over $\vec{x}$ in \Eq{eq:baryon3p} by an additional inversion. 
Then a generalized propagator for fixed time slice, projector $\mathbb{P}$
and sink momentum $\vp\pr$  is obtained, as indicated by the 
shaded area in the left diagram of  \Fig{fig:diagStandardAndStochastic}. 
This renders the explicit calculation of an all-to-all propagator unnecessary.

Our alternative method uses a stochastic estimate of the all-to-all propagator 
appearing in \threep functions like \Eq{eq:nucleonThreePointExample}. 
Such an estimate is obtained via
\begin{align}
 \label{eq:stochEstimate}
 \frac{1}{N_S} \sum\limits_{r=1}^{N_S} \eta_r^\dagger (\vec{y},\tau) 
 \phi(x,t)
  &\stackrel{N_S \to \infty}{\longrightarrow} \mathcal{M}^{-1}((\vec{x},t),(\vec{y},\tau)),\\
 \phi(x,t) &=\sum\limits_{\tilde{\vx}}\mathcal{M}^{-1}((\vec{x},t),\tilde{x})\ \eta_r(\tilde{x}) \nonumber
\end{align}
where $\mathcal{M}$ is the Dirac matrix. In the above equations we have suppressed Dirac and color indices. $\eta_r$ is a random source obeying 
\begin{align*}
  \frac{1}{N_S}\sum\limits_{r=1}^{N_S} \eta_{r,\text{a},\alpha}^\ast(x)\eta_{r,\text{b}, \beta} (y) 
  \stackrel{N_S \to \infty}{\longrightarrow} \delta_{xy}\delta_\text{ab} \delta_{\alpha\beta}\; .
\end{align*}
The stochastic method is diagrammatically illustrated in the right diagram of
\Fig{fig:diagStandardAndStochastic}.
Using the stochastic estimate, we can decompose the double sum in 
\Eq{eq:nucleonThreePointExample} into the product of two single sums, 
which is significantly computationally cheaper than the naive double sum and reads
\begin{multline*}
  C_3^{(p)}\left( t, \tau ; \vp, \vp\pr ; \mathbb{P} \right) =
  \mathbb{P}^{\alpha \alpha\pr}
  \sum\limits_{\vx}\: e^{-i\vp\pr\cdot\vx}\;
  \varepsilon_{cba}\: \varepsilon_{a\pr b\pr c\pr}\:
  \left( \mathcal{C} \gamma_5 \right)_{\beta\pr\gamma\pr}\:
  \left( \mathcal{C} \gamma_5 \right)_{\gamma\beta} \; \Gamma^{\delta\pr \delta} \delta_{d\pr d}
   \\\!\!\!\!\!\times
   \Big[
   S^{(u)}_{(\alpha\pr a\pr) (\alpha a)} (x, 0)\,
   S^{(u)}_{(\gamma\pr c\pr) (\gamma c)} (x, 0)\,
   \left( \eta(x) \gamma_5\right)_{\beta\pr  b\pr}\;
   \sum\limits_{\vy} e^{-i\vp\cdot\vy}\,
   S^{(d)}_{(\delta\pr d\pr) (\beta  b)} (y, 0)\,
   \left( \gamma_5 \phi^{\ast\,(d)} (y)\right)_{\delta d}
   \\
 - S^{(u)}_{(\alpha\pr a\pr) (\gamma c)} (x, 0)\,
   S^{(u)}_{(\gamma\pr c\pr) (\alpha a)} (x, 0)\,
   \left( \eta(x) \gamma_5\right)_{(\beta\pr  b\pr)}\;
   \sum\limits_{\vy} e^{-i\vp\cdot\vy}\,
   S^{(d)}_{(\delta\pr d\pr) (\beta  b)} (y, 0)\,
   \left( \gamma_5 \phi^{\ast\,(d)} (y)\right)_{\delta d}
   \Big]\,
\end{multline*}
where $x=(\vx,t)$ and $y=(\vy,\tau)$. We have suppressed the average over the number of stochastic samples, \cf \Eq{eq:stochEstimate}, 
and used the $\gamma_5$ Hermiticity property of the Dirac matrix to obtain the above expression. As before, we use superscripts to denote the quark flavor.

The drawback of the stochastic method is that we have to average over a sample of $N_S$ stochastic sources. 
This requires $N_S$ inversions compared to just twelve (one inversion per Dirac and color index) in the sequential method.
However, a major benefit of this method is its flexibility, since we do not need to fix the spin projector
or the sink momentum, nor even the sink time slice, in principle.

\section{\label{sec:test}Assessment of the stochastic method}
To test the applicability of the stochastic method, we need to determine 
how large $N_S$ must be in order to keep the stochastic noise
under control. 
This depends on the observable of interest.  To be concrete, we
compute a relatively simple benchmark observable of nucleon structure, 
namely the nucleon axial charge $g_A$,
using our stochastic method.
This quantity can be obtained from a nucleon matrix element of the isovector axial current.
For the evaluation of matrix elements of the nucleon, we need to introduce the zero momentum nucleon \twop function,
\begin{align*}
C_{2}^{(N)} (t) &=
\sum\limits_{\vx} 
\Tr\left[ \frac{1}{2} \left(\mathbf{1} + \gamma_0 \right) \left\langle \mathcal{I}^{(N)}(\vx,t) \right.
 \left. \bar{\mathcal{I}}^{(N)}(0) \right\rangle \right].
\end{align*}
We then examine the ratio $R_{g_A}$ of the nucleon \threep and \twop correlation functions,
\begin{align}
  \label{eq:ratio3pGA}
  R_{g_A}(t, \tau) &= \frac{C_{3,\Op_{k}}^{(N)} (t, \tau)}
                  {C_{2}^{(N)}(t)}  \\\nonumber
  C_{3,\Op_{k}}^{(N)} (t, \tau; \vp=\vp\pr=0) &= \sum\limits_{\vx,\vy} 
     \Tr\left[\Gamma_k \left\langle \mathcal{I}^{(N)}(\vx,t) \right| 
      \Op_{k}(\vy,\tau) 
     \left| \bar{\mathcal{I}}^{(N)}(0) \right\rangle \right]
   \\\nonumber
   \Op_{k}\left(\vy,\tau\right) &= \bar{u}(\vy,\tau) \gamma_5 \gamma_k u(\vy,\tau) - \bar{d}(\vy,\tau) \gamma_5 \gamma_k d(\vy,\tau), \\\nonumber
   \Gamma_k &= \frac{i}{2}\left(1+\gamma_0\right) \gamma_5 \gamma_k, \quad k=1,2,3 
\end{align}
In the limit of large Euclidean time separations, $R_{g_A}$ converges to $g_A$ up to the renormalization factor $Z_A$,
\begin{align*}
  Z_A R_{g_A}(t,\tau)  &\quad  \longrightarrow  \quad g_A 
  \quad \text{for} \quad 
    t\to \infty, \quad
    \tau\to \infty \quad \text{and}
  \quad (t-\tau)\to \infty.
\end{align*}
In this paper we use the value of the renormalization constant $Z_A=0.757(3)$ determined
non-perturbatively~\cite{Dimopoulos:2011wz, Alexandrou:2013joa}.

\begin{figure}[ht]
\includegraphics[width=0.8\textwidth]{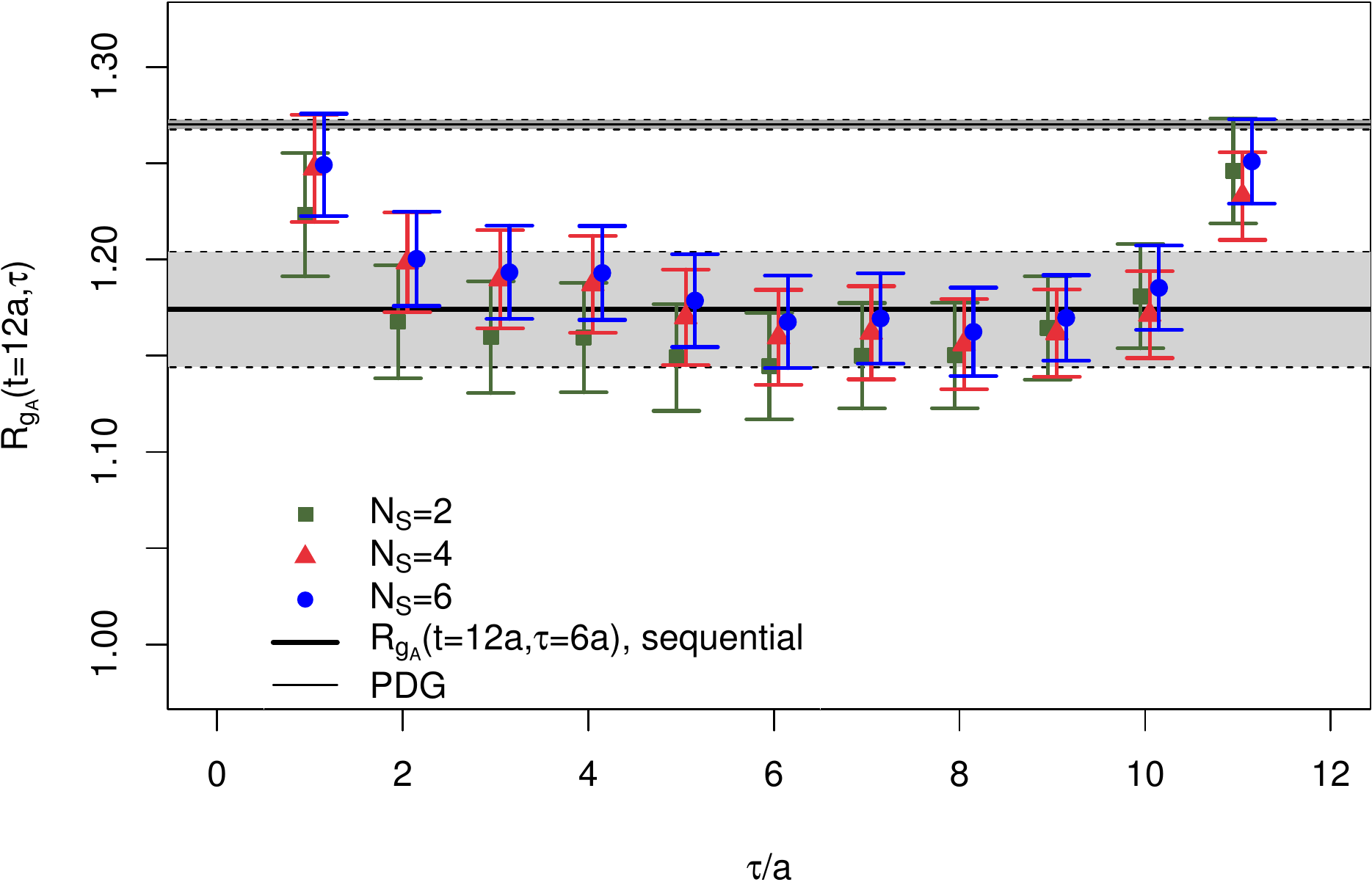}
\caption{\label{fig:gAstochastic} 
Plateau region of the ratio $R_{g_A}(t=12a,\tau)$ obtained from the 
stochastic method 
on one $N_f=2+1+1$ ensemble with a pion mass of $m_\pi\approx 370\mev$ and a lattice spacing $a \approx 0.082 \fm$. 
We use $N_S=2$, $4$ and $6$ spin-color diluted stochastic time-slice sources, with the source located at the sink.
The source-sink separation is $12a$ and we compare to the standard sequential method, 
of which we show the ratio $R_{g_A}(t=12a,\tau=6a)$ with the light gray band. 
The dark gray band indicates the PDG value\cite{Nakamura:2010zz}.}
\end{figure}

In order to demonstrate that the stochastic method can indeed produce results with a 
reasonable computational effort and is potentially competitive with the sequential method, 
we performed a benchmark calculation with $N_f=2+1+1$ flavors of quarks. We
employed twisted mass fermions at maximal twist 
with a lattice spacing of $a\approx 0.082~\text{fm}$ determined
from the nucleon mass~\cite{Alexandrou:2013joa}, a pion mass of $m_\pi\sim 370~\text{MeV}$
and a volume of $L^3\approx(2.6~\text{fm})^3$. 
In \Fig{fig:gAstochastic}, we show $R_{g_A}(t,\tau)$ obtained using the stochastic method
as a function of the insertion time $\tau$ for a fixed source-sink separation $t=12a$. 
We compare to the value $R_{g_A}(t=12a,\tau=6a)$ obtained using the sequential method. 
For different values of $\tau$ close to the middle of the plateau the picture is similar.
We use spin-color diluted random $Z(4)$ vectors as stochastic sources,
\begin{equation*}
 \eta^{a}_{\alpha} (\vx,t) = 
    \delta_{a,a_0} \delta_{\alpha,\alpha_0} \delta_{t,t_0} \tilde{\eta}(\vx), 
    \quad a_0 \in \{0,1,2\},\quad \alpha_0 \in \{0,1,2,3\}, \quad
 \tilde{\eta}(\vx) \in Z(4)\; .
\end{equation*}
We have used a fixed number of gauge field configurations $N_\text{gauge}=460$ in both our stochastic and sequential approaches.
Our observation is that using at most $N_S=4$ spin-color diluted stochastic noise vectors per
configuration for the estimate of the all-to-all propagator is sufficient to reach the same statistical accuracy as with the sequential method.

In terms of inversions, $N_S=1$ corresponds to using the same number of inversions as in the sequential method i.e. ($4 \times 12$) per gauge field configuration,
for every additional set of stochastic sources we need ($2 \times 12$) additional inversions to
obtain the forward and back propagators, respectively. 
Thus this method would require four times more inversions.
We would like to remark, however, that in the stochastic approach we can compute the correlation functions of proton and neutron
without additional inversions, in contrast to the sequential method. 
In addition we used $3$ operators $k=1,2,3$ in \Eq{eq:ratio3pGA},
and correspondingly $3$ spin projectors for the
stochastic method, again without the need of additional inversions when using the stochastic method.
Our observation is that this procedure reduces the statistical error by about a factor $\sqrt{6}$, corresponding
to a factor of about $6$ in statistics.
We would like to remark however that this effective gain in statistics 
is rather specific for $g_A$ and will change for other observables.

%
Having demonstrated that it is in fact possible to compute $g_A$ 
using the stochastic method with a reasonable computational effort
compared to the sequential method,
we would like to know how the situation changes when the volume is varied. 
A potential danger of our stochastic method is 
that the number of stochastic sources required to reach the same precision 
as the sequential method may increase rapidly as the volume increases.

To study the volume effects, 
we use $N_f=2$ flavors of maximally twisted mass fermions, instead of the $N_f=2+1+1$.  We expect that the
stochastic noise should not noticeably depend on the number of dynamical flavors, and for
the $N_f=2$ case, there exists a
series of four different volumes at the same value of the lattice spacing, $a\approx 0.082\fm$~\cite{Alexandrou:2010hf} 
and a pion mass, 
$m_\pi\approx 370\mev$~\cite{Boucaud:2008xu}. These volumes are $V=L^3\times T$, where $L/a=16, 20, 24, 32$ and the
temporal extent of the lattice is $T=2L$ in all cases. This enables us to thoroughly study the volume dependence of 
the stochastic method over a relatively large range of spatial 
volumes, from about $(1.4\fm)^3$ to $(2.8\fm)^3$.  This 
corresponds to $1.95 \lessapprox m_\pi L \lessapprox 3.9$.

We performed an analysis using the stochastic method on a fixed number of
gauge field configurations $N_\text{gauge}=200$ at each of the four 
volumes. The source-sink separation is fixed to $12a$ in all cases,
which corresponds to about $1.067$~fm.

\begin{figure}[htb]
\centering
\includegraphics[width=0.48\textwidth]{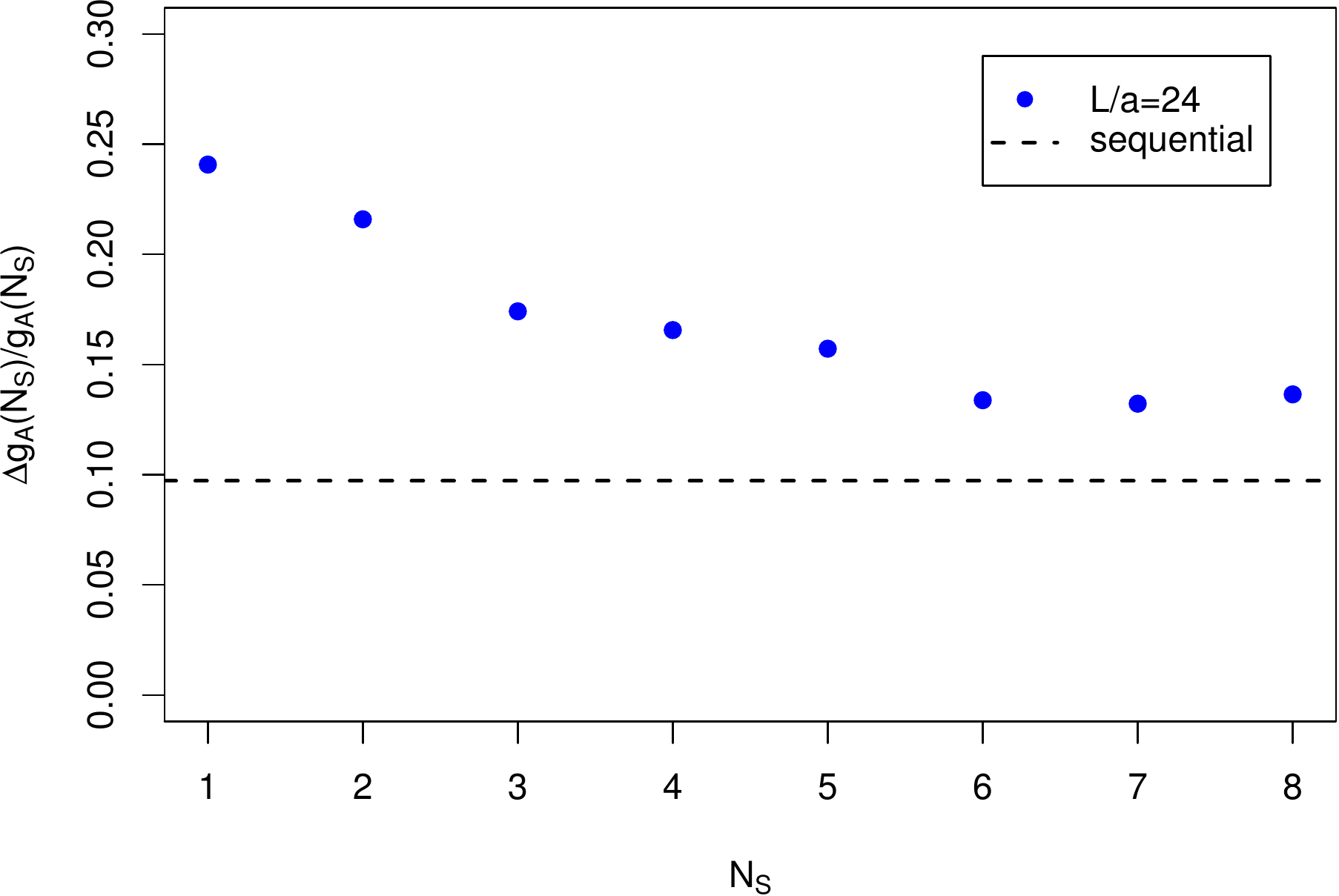}
\includegraphics[width=0.48\textwidth]{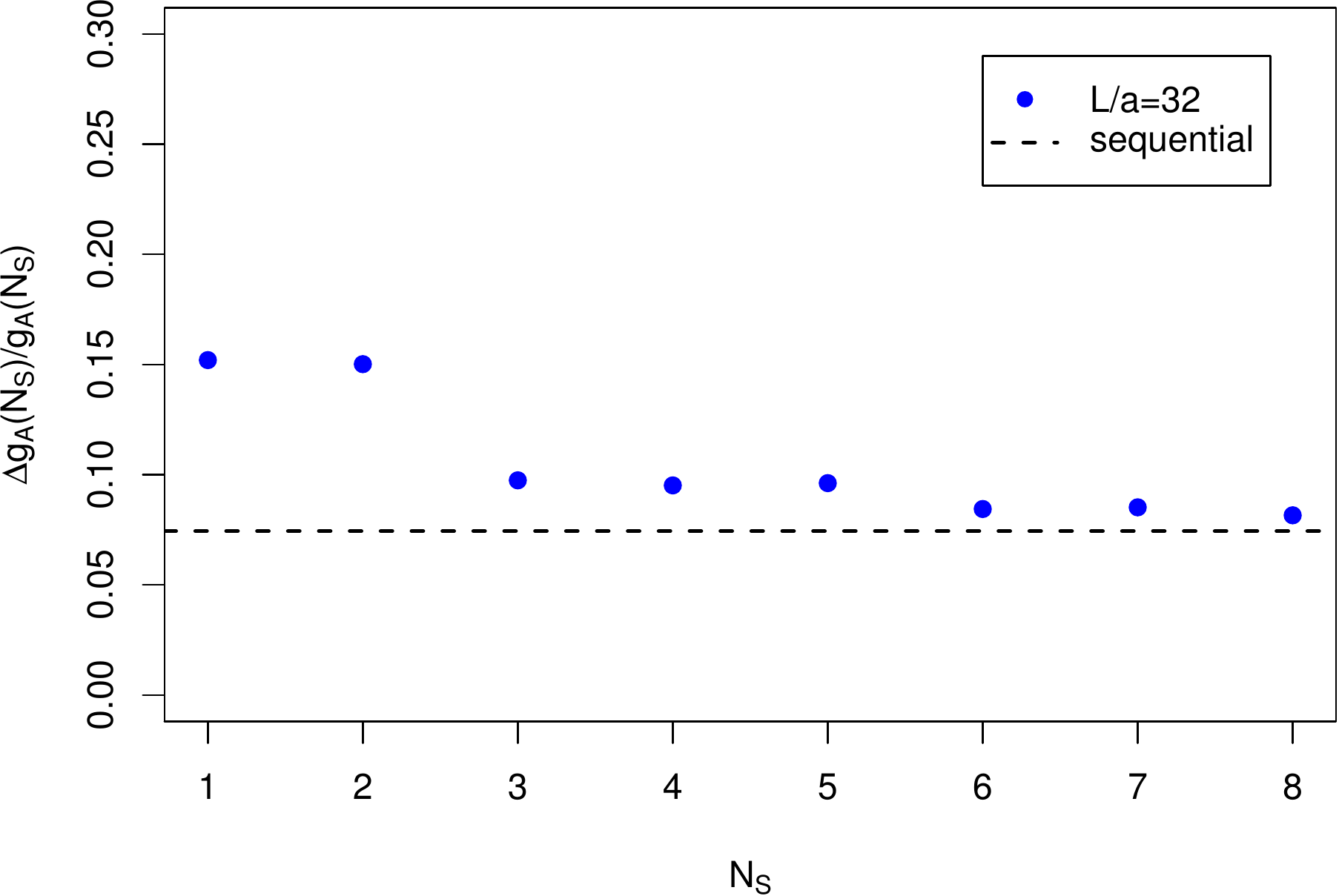}
\caption{\label{fig:stochMethodErrorConvergence}The filled circles show the 
relative error on the value obtained for $g_A$
as a function of the number of spin-color diluted stochastic sources $N_S$ per gauge field configuration. In the left panel
we show results for a lattice extent of $L/a=24$ and in the right panel we show results for $L/a=32$.
The dashed lines represent the error when using the sequential method.
A fixed number of gauge field configurations $N_\text{gauge}=200$ was used, 
with a pion mass of $m_\pi \approx 370\mev$ and a lattice spacing of $a\approx0.082\fm$.
}
\end{figure}
In \Fig{fig:stochMethodErrorConvergence} we show the relative error of $g_A$ as a function of
the number of stochastic sources for two of the four volumes $L/a=24$ and $L/a=32$,
where also the sequential method has been applied.
For both volumes a convergence towards the error of the sequential method can be seen,
which looks better for the larger volume, where the error is close to the error of the
sequential method for $N_S=4$, to be conservative. This confirms the observation in the $N_f=2+1+1$ calculation mentioned above, which was done
at about the same physical volume. The error of the error is not shown, but it is roughly of order $10\%$ of the error. 

\begin{figure}[htb]
\centering
\includegraphics[width=0.7\textwidth]{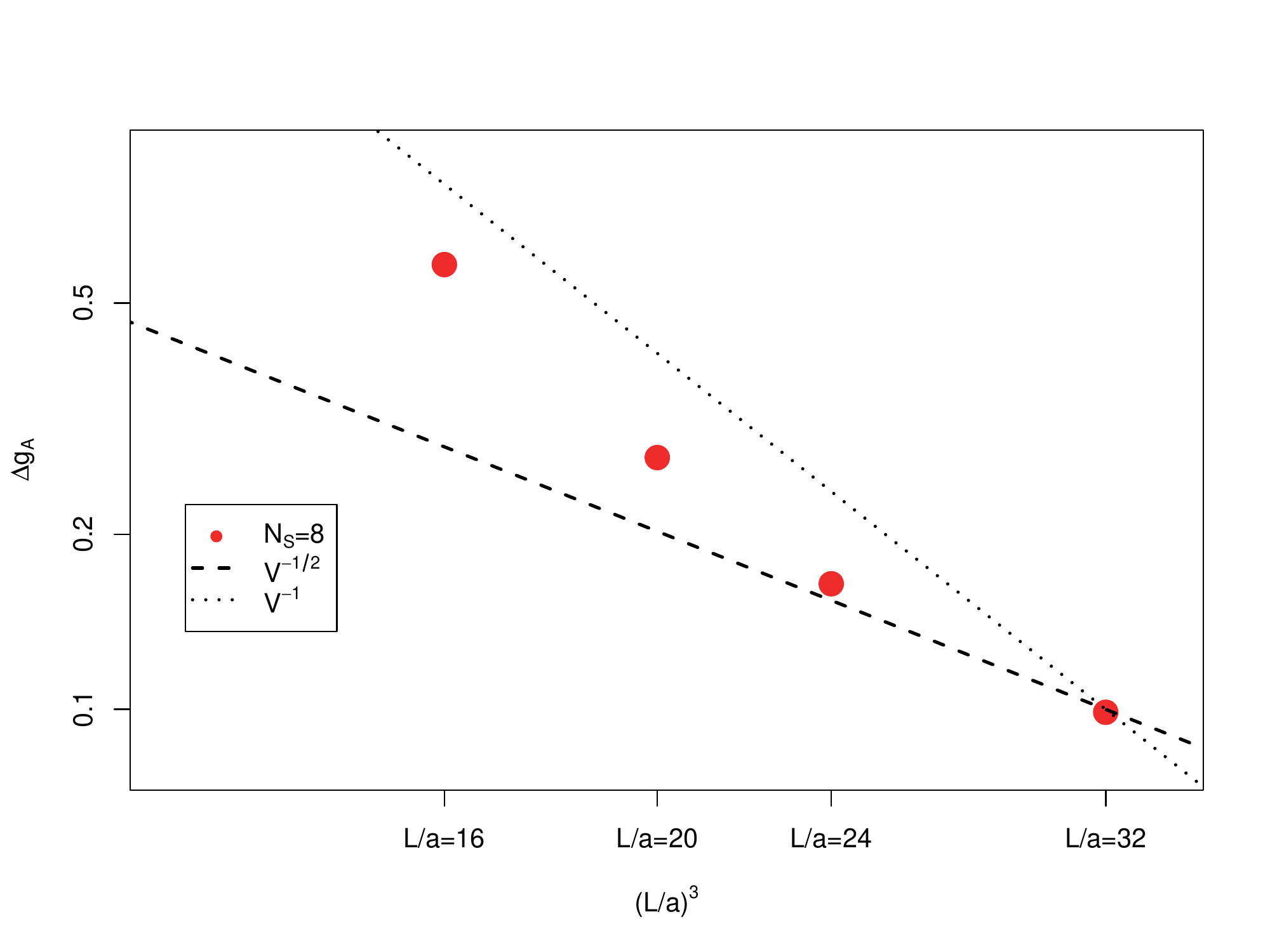}
\caption{\label{fig:stochMethodVolumeScaling}The filled circles
show the error of $g_A$ using the stochastic method with
a fixed number of stochastic sources $N_S=8$
on a fixed number of gauge fields $N_\text{gauge}=200$ for four different lattice sizes. 
The pion mass is $m_\pi \approx 370\mev$ and the lattice spacing is $a\approx0.082\fm$.
The dashed lines indicate a scaling proportional to $V^{-1/2}$ and $V^{-1}$ 
and are solely meant to guide the eye. The error of the error is not much bigger than the symbol size, hence we do not show it.
Please note the double logarithmic scale.}
\end{figure}

We show the combined statistical and stochastic error of $g_A$ obtained using the stochastic 
method with $N_S=8$ for four different volumes 
in \Fig{fig:stochMethodVolumeScaling}.
In order not to be subject to a systematic error we 
do not fit the ratio given in \Eq{eq:ratio3pGA} using an 
estimated plateau range $\tau$ between source and sink.
Instead, we take the renormalized ratio $R_{g_A}(t=12a,\tau=6a)$ and its error in 
the middle between source and sink as our estimate for $g_A$, 
where contributions from excited states are expected to be the smallest.
For the larger volumes the error scales like the one of the sequential method, $V^{-0.5}$~\cite{Beane:2009gs}.
Therefore the plot is consistent with the error being dominated by the gauge noise for larger volumes, 
which we have demonstrated above, see \Fig{fig:stochMethodErrorConvergence}.
Thus this method appears to work as well or even better at the larger volumes 
typically employed in current calculations, an observation which, of course, 
needs to be verified for other quantities and different physical situations.

\begin{figure}[htb]
\centering
\includegraphics[width=0.49\textwidth]{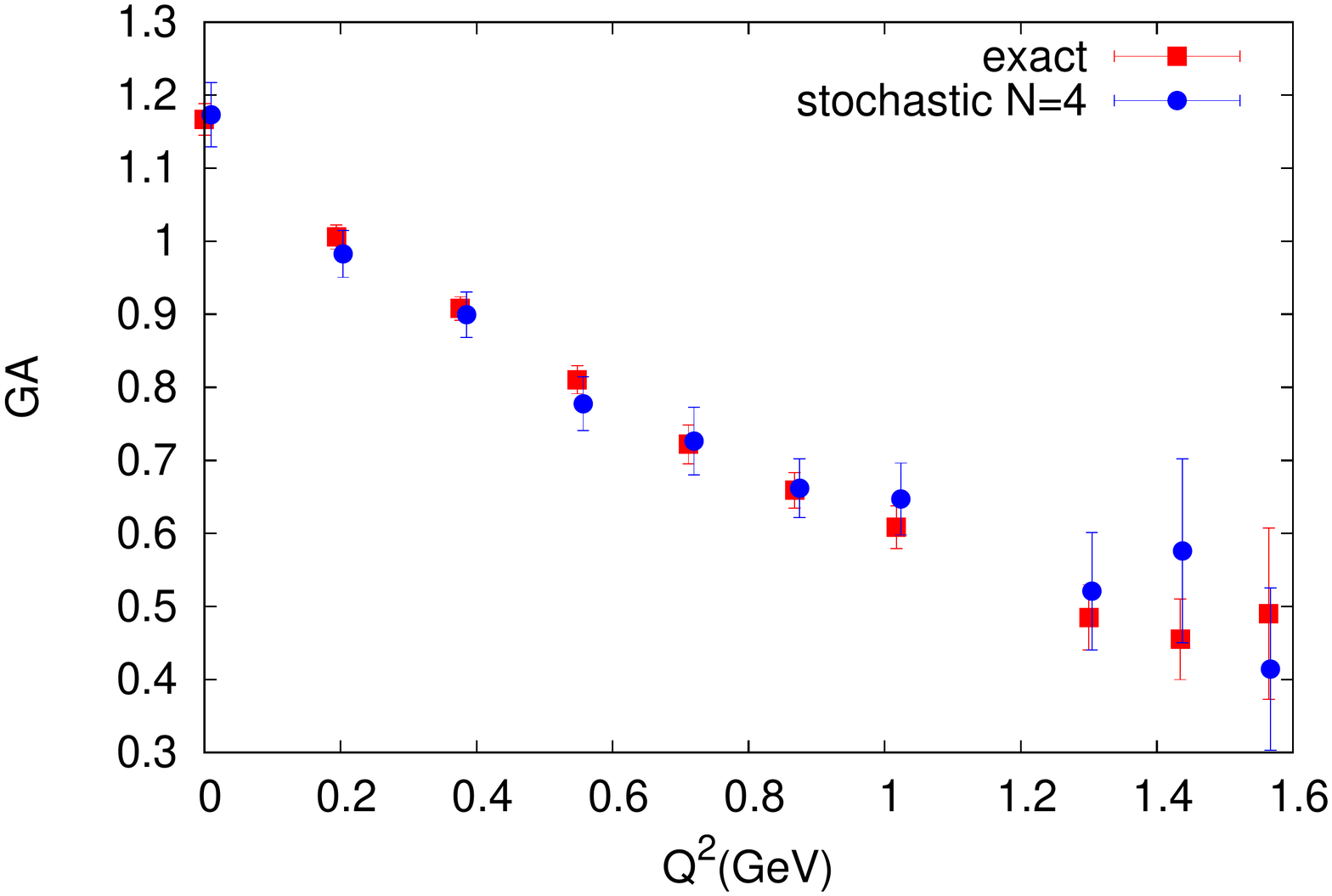}
\includegraphics[width=0.49\textwidth]{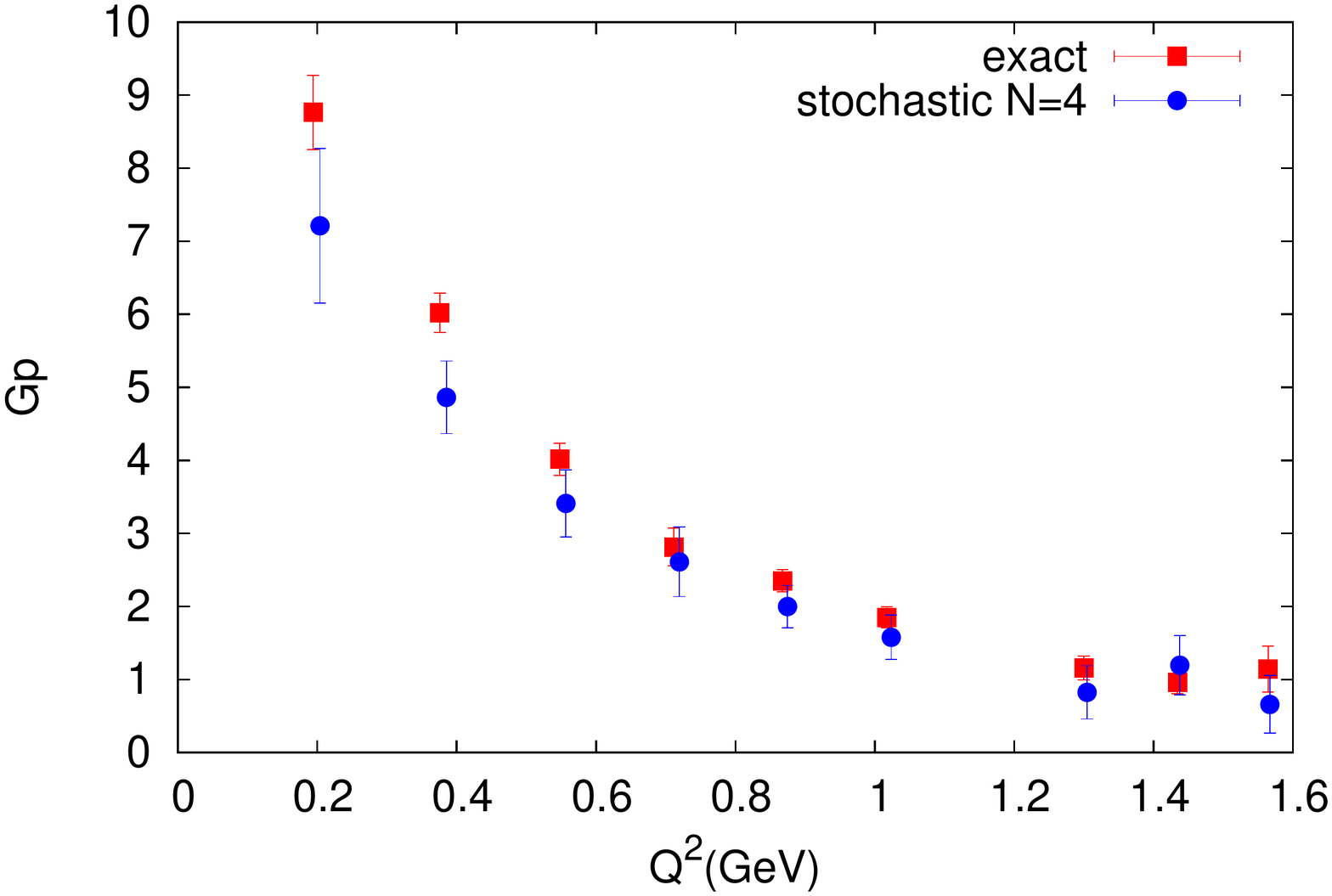}
\caption{\label{fig:GA_GP} Comparison of $G_A(Q^2)$ and $G_p(Q^2)$ computed  with the stochastic and standard sequential method (refered as exact in the legend). For the stochastic estimate we use a fixed number of stochastic sources $N_S=4$ and we use one projector for the proton correlators. The computation is performed  using $N_f=2+1+1$ configurations fixed to $N_\text{gauge}=500$ with a pion mass of $m_\pi \approx 370\mev$ and a lattice spacing of $a\approx0.082\fm$.}
\label{fig:GA Gp}
\end{figure}

The suitability of the method is demonstrated in Fig.~\ref{fig:GA Gp} where
we compare the nucleon axial form factors $G_A(Q^2)$ and $G_p(Q^2)$~\cite{Alexandrou:2013joa} computed using the standard approach and our new method. For this comparison we use the same number of configurations and one projector. Since we use four stochatic noises diluted in colour and spin the computation is four times more expensive producing errors that are comparable to the exact case.
However, the new method allows to compute the three point functions of the
neutron as well as for three different projectors for free thus compensating for the increases cost. This compared with the fact that one can consider different  final states e.g. a nucleon carrying momentum for free makes the new method more versatile.  
%
%
%
\section{\label{sec:sum}Conclusion}
We have applied a stochastic method for the 
calculation of nucleon matrix elements using spin-color diluted time slice sources. We have taken the case of the nucleon axial charge as a typical example
of a three-point function to explore the method.
Our conclusions for our test case is that the error is comparable to 
the error of the sequential method already at a moderate number of stochastic sources, 
namely four spin and color diluted timeslice stochastic sources, 
when using the same number of gauge field configurations. 
In this particular case we have effectively increased
the statistics used in the stochastic method by a factor $6$ through averaging over 
neutron and proton correlators and using $3$ different currents, 
which does not require the computation of new propagators.
Moreover, our results indicate that the convergence behaviour in the number 
of stochastic sources, $N_S$, appears to improve when the volume is increased.

Since the stochastic method needs of  ${\cal O}(10)$ more inversions 
(we needed $N_S=4$ for $g_A$ but this maybe different for other matrix elements) 
 it is competitive with the
sequential method when one computes ${\cal O}(10)$ more types of
matrix elements. In the case of $g_A$ the increase in computational effort
is easily compensated by computing 6 different matrix elements with 3 
different spin projections for the proton and the neutron.  
Thus, even with the additional computational overhead, the great versatility of the stochastic 
method explained in this paper outweighs the sequential method when many baryon matrix elements 
or form factors are computed.  

\section*{Acknowledgments}
We thank our fellow members of ETMC for their constant collaboration.
We are grateful to the John von Neumann Institute for Computing (NIC),
the J{\"u}lich Supercomputing Center and the DESY Zeuthen Computing
Center for their computing resources and support. This work has been
supported in part by the DFG Sonderforschungsbereich/Transregio
SFB/TR9.  It has been coauthored by Jefferson Science Associates, LLC
under Contract No.\ DE-AC05-06OR23177 with the U.S.\ Department of Energy.
This work is supported in part by the Cyprus Research Promotion
Foundation under contracts KY-$\Gamma$/0310/02/, and the Research Executive Agency of the European Union under Grant Agreement number PITN-GA-2009-238353 (ITN STRONGnet).
K. J. was supported in part by the Cyprus Research Promotion
Foundation under contract $\Pi$PO$\Sigma$E$\Lambda$KY$\Sigma$H/EM$\Pi$EIPO$\Sigma$/0311/16.

%
%
\bibliography{stochastic_3point}
\end{document}